%% file: main.tex
\documentclass[aps,prl,twocolumn,floats,balancelastpage,showpacs,showkeys,preprintnumbers,floatfix,nofootinbib,superscriptaddress, longbibliography]{revtex4-1}
\usepackage{graphicx,amsmath,amssymb,amsfonts, soul, amssymb,color,float,wasysym,wrapfig,xspace,changepage,todonotes}
\usepackage[colorlinks]{hyperref}
\usepackage[utf8]{inputenc}

\DeclareUnicodeCharacter{0327}{\\c{}} 
\DeclareUnicodeCharacter{2212}{-} 

\usepackage{orcidlink}
\usepackage{lineno}

\begin{document}
\title{Search for gamma-ray spectral lines from dark matter annihilation with the H.E.S.S. Inner Galaxy Survey}

\input{authors_prl}

\begin{abstract}
Spectral gamma-ray line features are expected as key signatures from dark matter (DM) annihilations of TeV-scale particle DM. Observations of the Galactic Centre with atmospheric Cherenkov telescopes are unique to probe thermal-relic TeV particle DM, well beyond the reach of direct detection and collider searches. We report here on the search for line signals in very-high-energy gamma rays using data from the Inner Galaxy Survey, consisting of 546 hours of H.E.S.S. observations of the inner few degrees of the Galactic Centre. No significant signal is detected. We then compute the exclusion limits on the annihilation line cross section $\langle \sigma v \rangle_{\rm line}$, with a two-dimensional log-likelihood ratio test statistics, exploiting spectral and spatial features of the DM signal. Assuming an Einasto DM density profile for the Milky Way, our results provide the most constraining limits so far, reaching $\langle \sigma v \rangle_{\rm line} = 2.3$ $\times$ $10^{-28}$ and $2.4  \times$ $10^{-27}$ cm$^3$s$^{-1}$ for DM masses of 1 and 10 TeV, respectively. The present limits are used to constrain the widely searched Wino, Higgsino and Quintuplet models. For the first time, thermal Higgsino DM is probed for DM Milky Way models.
\end{abstract}


\maketitle

\section{Introduction}
\label{sec:introduction} 
Cosmological and astrophysical measurements show that about 85\% of the matter content of today's Universe is made of non-baryonic cold dark matter (DM), \textit{e.g.}~\cite{Planck:2018vyg}. The properties of dark matter, beyond what can be inferred from its gravitational interaction, remain 
a forefront open question of fundamental physics. Long recognised for its simplicity and elegance, the model of Weakly Interacting Massive Particles (WIMP) provides archetypal elementary particle DM candidates: with masses in the GeV -- 100 TeV mass range, they naturally explain~\cite{Bergstrom:2000pn} the cold DM density measured today, if thermally-produced in the early Universe. WIMP realisations naturally arise from either a full theory beyond the Standard Model, \textit{e.g.}, supersymmetry~\cite{Jungman:1995df}, or its minimal extensions~\cite{Cirelli:2005uq,Cirelli:2007xd,Cirelli:2008id,Cirelli:2009uv,Cirelli:2015bda,Mahbubani:2005pt,Kearney:2016rng}. 

Searches 
with direct detection experiments
and colliders~\cite{Schumann:2019eaa,Kahlhoefer:2017dnp} have begun to exclude many WIMP-scenario realisations, in particular for masses up to the 100 GeV scale. When reaching TeV masses, some long-predicted and actively searched models such as the Wino and Higgsino remain completely unconstrained and out of reach of future detection prospects at the LHC~\cite{Bottaro:2021snn,Bottaro:2022one}: searches have ruled out Wino masses $\lesssim$500 GeV; the Higgsino is considerably more difficult to search for, even $\sim$400 GeV is hardly probed with the full LHC dataset. 
The thermal Higgsino ($m \simeq$ 1 TeV) detection is so challenging that it would be only within the reach of the proposed long-term future colliders, such as a 10 TeV muon collider~\cite{Capdevilla:2021fmj} or a 100 TeV future circular hadron collider (FCC-pp)~\cite{Saito:2019rtg}. 
For direct detection,
the thermal Wino scattering cross section is near the neutrino floor, and thermal Higgsino DM is out of reach of ongoing or planned underground experiments as its scattering cross section is well below the neutrino floor, \textit{e.g.}~\cite{Hill:2011be,Hill:2013hoa,Hill:2014yka,Hisano:2015rsa,Chen:2019gtm}. Virtually all TeV-mass WIMP annihilation models are predicted to provide gamma rays in the final state~\cite{Cirelli:2010xx}.
Besides a continuum of gamma-rays produced by hadronization and/or decays as a consequence of annihilation products, TeV-mass WIMP models exhibit prominent line-like features at energies close to their mass $m_{\rm DM}$ in their self-annihilation energy spectrum, such as in the self-annihilation process at rest into $\gamma X$ with $X = \gamma, h, Z$ or a non-Standard Model neutral particle, providing a spectral line at an energy $E_\gamma = m_{\rm DM}(1-m^2_{\rm X}/4m^2_{\rm DM})$, limited only by the detector resolution given the low ($\sim$10$^{−3}$ c) relative velocity of the DM particles.

Minimal DM models add only a minimal field content to the Standard Model, rather than aiming for a UV-complete WIMP theory, \textit{e.g.}, Supersymmetry, providing a full description of the particles and their interactions valid at all energy scales. Such models consider TeV-scale states charged under the electroweak interaction including a SU(2) doublet with unit hypercharge, and a 3 and 5 representation of SU(2), dubbed as Higgsino, Wino, and Quintuplet, respectively. They produce the correct DM abundance when embedded in a thermal relic cosmology, if their masses are 1.0~$\pm$~0.1~TeV, 2.9~$\pm$~0.1~TeV, and 13.6~$\pm$~0.8 TeV, respectively~\cite{Cirelli:2007xd,Hisano:2006nn,Hryczuk:2010zi,Beneke:2016ync,Mitridate:2017izz,Bottaro:2021snn}. Interestingly, the former two naturally emerge in supersymmetry. For each of these WIMP realisations, the full determination of the cross section and gamma-ray yield per annihilation requires the following processes to be included: the Sommerfeld enhancement, resummation of effects of the order of $m_{\rm DM}/m_{\rm W}$, and additional channels beyond the direct annihilation to two photons, including the part of the spectra accounting for end-point photons and continuum emission, and the production and decay of bound states, \textit{e.g.}~\cite{Baumgart:2014vma,Baumgart:2015bpa,Baumgart:2018yed,Beneke:2018ssm,Beneke:2019gtg,Beneke:2019vhz,Rinchiuso:2020skh,Beneke:2022eci,Baumgart:2023pwn}. For such models, the annihilation cross section into photon pairs is largely enhanced compared to the $1/\alpha^2$ loop-suppressed value.  

The search for gamma rays from DM annihilation focuses on regions of the sky close to the observer and with large DM content, such as the Galactic Centre (GC). Nearby dwarf galaxies are alternative compelling targets, however the DM signals are expected to be lower than from the GC~\cite{Strigari:2018utn}. 
The GC region exhibits considerable conventional gamma-ray emission from astrophysical sources and cosmic-ray interactions with gas and ambient photons. These challenge searches for DM. Besides the spatial distribution of the expected DM signal that scales by the square of the DM density, line-like signals provide a powerful discriminating feature against the much smoother background, which cannot mimic such a spectral shape. 

Given its location in the Southern Hemisphere (23°16'17''S, 16°30'00''E), the H.E.S.S. observatory of Imaging Atmospheric Cherenkov Telescopes (IACTs) is ideally situated to observe the GC
under optimal conditions. H.E.S.S. can observe the GC for more than 300 hours per year at zenith angles below 30$^\circ$~\textbf{\cite{realobs}}. The strongest limits so far on the DM line cross section are obtained with 254 hours of H.E.S.S. GC observations 
up to masses of 20 TeV~\cite{Abdallah:2018qtu}, and 223 hours of large-zenith-angle observations by MAGIC for masses beyond 20 TeV~\cite{MAGIC:2022acl}. In this Letter, we present new constraints from 546 hours of H.E.S.S. observations, obtained with the Inner Galaxy Survey (IGS) program. The limits significantly improve over previous ones in the 300 GeV -- 60 TeV mass range, 
pushing the sensitivity to Higgsino, Wino and Quintuplet DM models even further. For the first time, thermal Higgsino DM can be probed.

\section{Expected gamma-ray flux} 
\label{sec:flux} 
The differential gamma-ray flux from self-annihilating Majorana DM particles of mass $m_{\rm DM}$ into two photons, in a solid angle $\Delta\Omega$, writes as:
\begin{equation}
\begin{aligned}
\label{eq:promptflux}
\frac{{\rm d} \Phi}{{\rm d} E_{\gamma}} (E_{\gamma}, \Delta\Omega) = \frac{\langle \sigma v \rangle_{\rm line}}{8\pi \ m_{\rm DM}^2} 
 \frac{{\rm d} N_\gamma}{{\rm d} E_{\gamma}}(E_{\gamma})\times J(\Delta\Omega) \ ,  \\
{\rm with} \, \, \, \,  J(\Delta\Omega)=\int_{\Delta\Omega}\int_{\rm LOS}{\rm d} s \ {\rm d} \Omega \ \rho^2(r(s,\theta))\ ; 
 \end{aligned} 
\end{equation} 
where $\langle \sigma v\rangle_{\rm line}$ is the velocity-weighted line annihilation cross section averaged over the velocity distribution of DM, and $dN_\gamma/dE_\gamma(E_\gamma) = 2 \delta$($E_\gamma - m_{\rm DM}$) is the differential yield of gamma rays per annihilation. $J(\Delta\Omega)$ encapsulates the squared DM density integrated over the line of sight (LOS) and the solid angle $\Delta\Omega$.  The coordinate $r$ is given by $r = (r^2_{\odot}+s^2-2 r_{\odot} s\ {\rm cos}\ \theta)^{1/2}$, where $s$ is the distance along the line of sight, and $\theta$ the angle between the direction of observation and the GC. $r_{\odot}$ is the distance 
between the observer 
and the GC, taken as $r_{\odot}$ = 8.178~kpc~\cite{GRAVITY:2018ofz}. For TeV electroweak candidates such as the Wino, Higgsino and Quintuplet, the total cross section is dominated by the line cross section which includes the contributions of the two-body processes $\gamma\gamma$ and $\gamma Z$, as well as the endpoint contribution arising from three-body or more final states containing one photon with energy very close to the DM mass, \textit{i.e.}, $1-E_\gamma/m_{\rm DM}\ll1$. Since we consider 
heavy WIMPs, we neglect the fact that the finite Z boson mass causes $E_\gamma = m_{\rm DM}(1-m^2_{\rm Z}/4m^2_{\rm DM})$. The  IACT energy resolution of $\sim$10\% cannot separate the $\gamma\gamma$ and $\gamma Z$ signals for TeV DM mass, such that  $\langle\sigma v \rangle_{\rm line} = \langle\sigma v \rangle_{\gamma\gamma} + 1/2 \langle\sigma v \rangle_{\gamma Z}$, as there is only a single photon in the latter process. In addition, with such an energy resolution, endpoint photons
contributing to $\langle\sigma v \rangle_{\rm line}$ cannot be distinguished from those produced in the
$\gamma\gamma$ and $\gamma Z$ final states~\cite{Beneke:2022eci,Baumgart:2023pwn}. Constraints derived from the search of the line signal can therefore be compared to the predicted line cross section for the aforementioned canonical models (see for instance, Refs.~\cite{Baumgart:2017nsr,Rinchiuso:2020skh,Beneke:2022eci,Montanari:2022buj}).

The J-factors expected in the GC region are subject to considerable uncertainties. The determination of the DM density in the central kpc of the Galaxy can be obtained either from mass modelling using kinematic measurements of the gravitational potential~\cite{Portail:2016vei,Cautun:2019eaf}, cosmological simulations of structure formation of Milky Way-like galaxies including only DM ~\cite{Navarro:1996gj,Springel:2008by}, or also accounting for baryonic physics and feedback processes~\cite{Hopkins:2017ycn,Board:2021bwj,McKeown:2021sob}. In this work, we use the Einasto profile as a baseline~\cite{Springel:2008by} with solar distance value of $r_{\odot}$~=~8.178~kpc, and explore the impact of changing $r_{\odot}$ to 8.5~kpc~\cite{Ghez:2008ms}. To bracket the uncertainty on the DM density models, we adopt the Navarro-Frenk-White (NFW) profile and its contracted version (cNFW) obtained from modelling of the MW mass profile using Gaia DR2 measurements~\cite{Cautun:2019eaf,Montanari:2022buj,supplement}, and the non-parametric FIRE-2~\cite{McKeown:2021sob} and Auriga~\cite{Hussein:2025xwm} profiles computed from hydrodynamical simulations of MW-like galaxies. The Einasto and NFW DM profiles are normalised to the local DM density $\rho_\odot$, such that $\rho$($r_\odot$) = $\rho_\odot = 0.39$ GeV cm$^{-3}$~\cite{Catena:2009mf}. Nevertheless, an improved determination of the local DM density would affect the overall DM distribution. For the Einasto profile, the integrated J-factor over the region for the DM signal search is 2.5$\times$10$^{22}$ GeV$^2$cm$^{-5}$~\cite{supplement}.

\section{Observations and Data analysis} 
\label{sec:analysis} 
This work makes use of the data collected with the five-telescope H.E.S.S. array between 2014 and 2020, for a total live time of 546 hours and an average zenith angle of 18$^\circ$. This dataset was first used in Ref.~\cite{HESS:2022ygk}.

The technique detailed in Ref.~\cite{deNaurois:2009ud} was applied to select and reconstruct gamma-ray events by fitting the shower images to semi-analytical models.
This yields an angular resolution of $0.06^\circ$ (68\% containment radius) and an energy resolution of 10\% above 300 GeV. The line signal is searched for in regions of interest (ROIs) defined as concentric annuli of 0.1$^\circ$ width, centred on the GC position, and with inner radii ranging from 0.5$^\circ$ to 2.9$^\circ$. The regions used for the background measurement (OFF) are built symmetrically to the ROIs from the pointing position, used for signal measurement (ON). A conservative set of masks is applied to avoid the challenging modelling of gamma-ray contamination from known VHE astrophysical sources in the centre of the Milky Way. More details are provided in Ref.~\cite{supplement}.

Following Ref.~\cite{HESS:2022ygk}, the statistical data analysis is based on a two-dimensional log-likelihood ratio test statistics (TS).
A spectral line at $E_\gamma = m_{\rm DM}$, folded with the H.E.S.S. energy resolution, is searched in the 61 logarithmically-spaced energy bins with central values from 300 GeV up to 64 TeV and 25 spatial bins corresponding to the ROI annuli. The 2-dimensional likelihood function combines the statistically independent measurements in the ON and OFF regions, expressed in the $i$th energy and $j$th spatial bins by:
\begin{widetext}
\begin{equation}
\mathcal{L}_{\rm ij}({\bf N}^{\rm S},{\bf N}^{\rm B}|{\bf N}_{\rm ON},{\bf N}_{\rm OFF}) = \frac{[\beta_{\rm ij}(N_{\rm ij}^{\rm S}+N_{\rm ij}^{\rm B})]^{N_{{\rm ON, ij}}}}{N_{{\rm ON, ij}}!}e^{-\beta_{\rm ij}(N_{\rm ij}^{\rm S}+ N_{\rm ij}^{\rm B})} 
\frac{[\beta_{\rm ij}(N_{\rm ij}^{\rm S'}+N_{\rm ij}^{\rm B})]^{N_{{\rm OFF, ij}}}}{N_{{\rm OFF, ij}}!}e^{-\beta_{\rm ij}(N_{\rm ij}^{\rm S'}+N_{\rm ij}^{\rm B})} e^{-\frac{(1-\beta_{\rm ij})^2}{2\sigma_{\beta_{\rm ij}}^2}} \, .
\label{eq:lik}
\end{equation}
\end{widetext}
The measured number of events in the signal and background regions is indicated by $N_{\rm ON,ij}$ and $N_{\rm OFF,ij}$.
The expected number of background events in both regions is given by $N^{\rm B}_{\rm ij}$.
$N^{\rm S}_{\rm ij}$ and $N^{\rm S'}_{\rm ij}$ represent the number of events expected from DM self-annihilation in the ON and OFF regions, respectively.
These are obtained by folding the DM flux in Eq.~(\ref{eq:promptflux}) with the H.E.S.S. energy-dependent acceptance and energy resolution, the latter conservatively assumed to be a Gaussian function of $\sigma_E/E$ of 10\% above 300~GeV \cite{deNaurois:2009ud}. 
The energy-dependent acceptance is computed on a run-by-run basis with run-wise simulations of the instrument response functions (IRFs)~\cite{Holler:2020duc}. This represents the most accurate and realistic IRF determination possible to date for H.E.S.S. data analysis, including observational conditions and detector configurations, and it allows for an improved control of the instrumental systematic uncertainty~\cite{Holler:2020duc}. The J-factors per ROI are reported in Tab. II of Ref.~\cite{supplement}, for each MW DM model used in this work. Following Refs. ~\cite{Silverwood:2014yza,Lefranc:2015pza,Moulin:2019oyc,HESS:2022ygk,Montanari:2022buj}, the observational and instrumental systematic uncertainties can be accounted for in the likelihood function with a Gaussian nuisance parameter. The normalisation factor for the expected number of events is expressed by $\beta_{\rm ij}$, and $\sigma_{\beta_{\rm ij}}$ is the width of the Gaussian function. The $\beta_{\rm ij}$ parameter is computed by maximizing the likelihood function for $\rm d\mathcal{L}_{\rm ij}/d \beta_{\rm ij} \equiv 0$. $\sigma_{\beta_{\rm ij}}$ is fixed to a flat value of 1\%, as determined in Ref.~\cite{HESS:2022ygk}. Using a Poisson term for the signal and background regions, respectively, in the likelihood function properly accounts for the impact of the background fluctuations in the TS computation, and therefore in the limits. Exclusion regions allow for the minimisation of artefacts in the background estimates due to leakage of VHE emission from known sources. Moreover, measuring the residual background allows us to avoid modelling the expected background or fitting it to the data.

In the absence of a significant line-signal excess, constraints on $\langle \sigma v\rangle_{\rm line}$ are computed from the TS, assuming a positive signal $\langle \sigma v\rangle_{\rm line} >$ 0~\cite{2011EPJC711554C}.
We adopted the high statistics limit in which the TS follows a $\chi^2$ distribution with one degree of freedom. Values of $\langle \sigma v\rangle_{\rm line}$ with a corresponding TS higher than 2.71 are excluded at the 95\% confidence level (C.L.). Expected limits are also computed from 300 independent Poisson realisations of the expected background in the ON and OFF regions, respectively. From the distribution of these values, the mean and the 68\% and 95\% containment bands are extracted. The means of the distributions provide 
the mean expected upper limits at 95\% C.L.
\begin{figure*}[htbp!]
\includegraphics[width=0.44\textwidth]{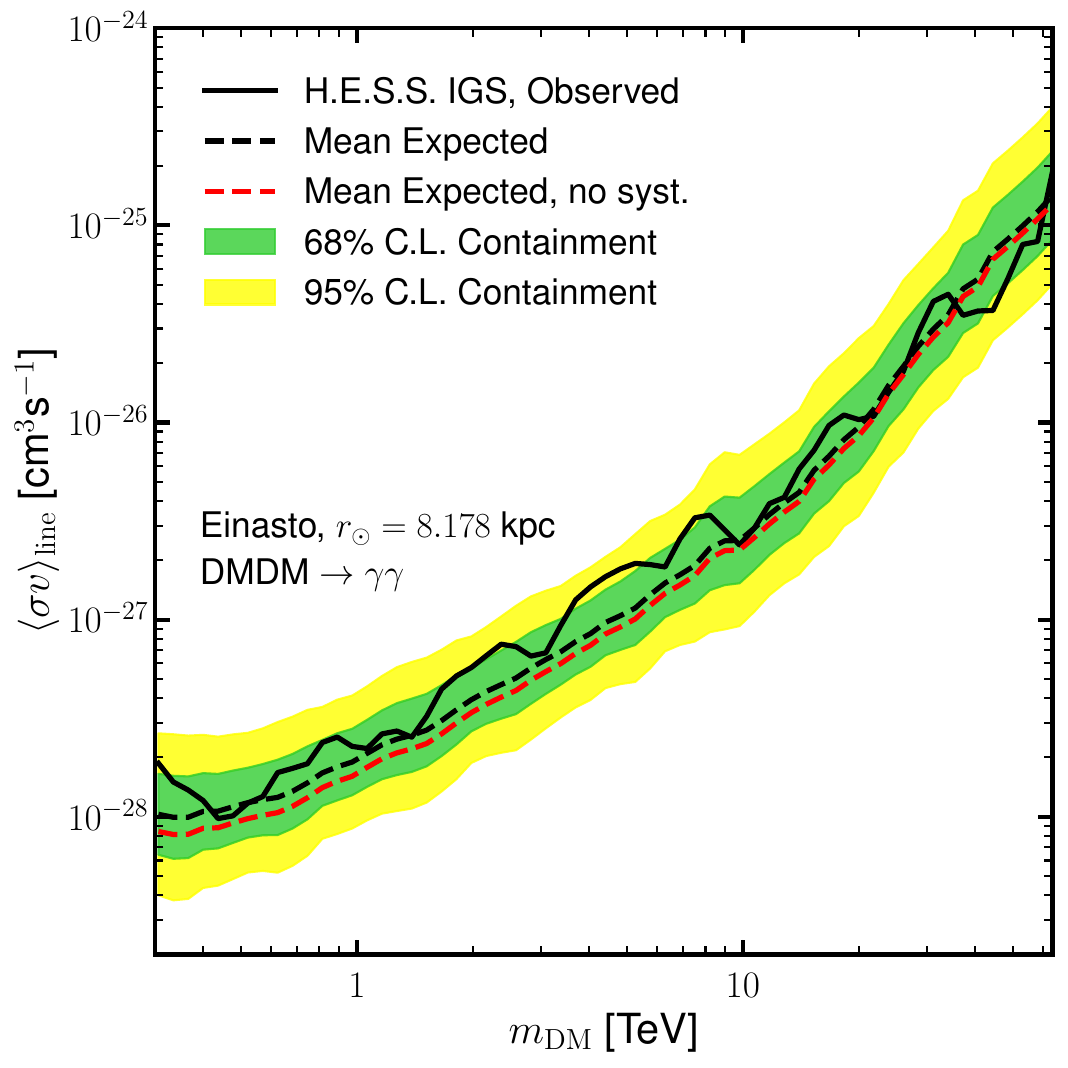}
\hspace{0.5cm}
\includegraphics[width=0.44\textwidth]{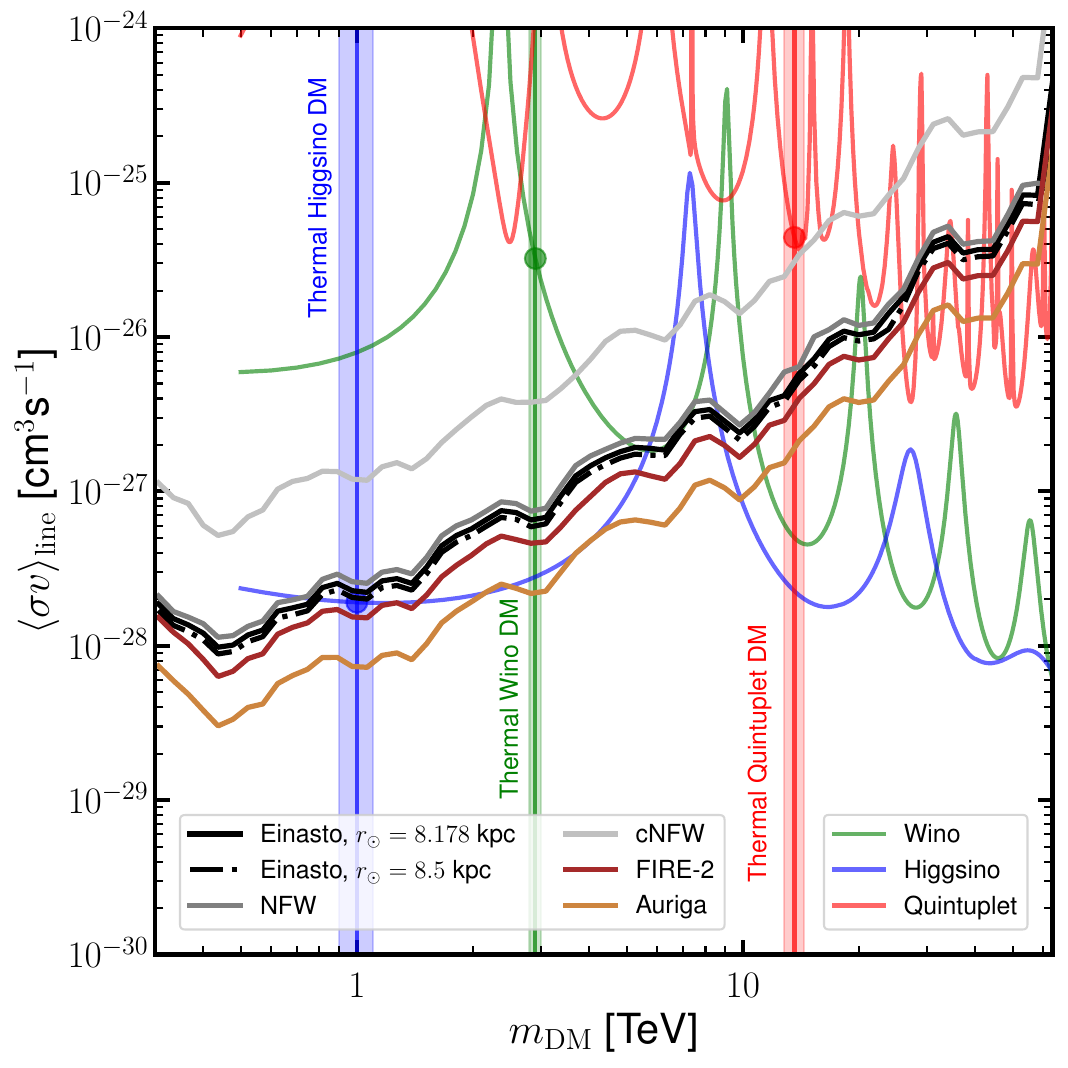}
\caption{
{\it Left panel:} 95\% C. L. observed upper limits (solid black line) on the line cross section $\langle \sigma v \rangle_{\rm line}$ versus the DM mass $m_{\rm DM}$ assuming the baseline Einasto profile for the DM distribution. 95\% C. L. mean expected upper limits (black dashed line) together with the 1$\sigma$ (green shaded area) and 2$\sigma$ (yellow shaded area) containment bands, respectively, are also plotted. All these upper limits include systematic uncertainty. The mean expected upper limit without systematic uncertainty is drawn as the red dashed line. \textit{Right panel:} 95\% C. L. observed upper limits on the line cross section $\langle \sigma v \rangle_{\rm line}$, where $\langle\sigma v \rangle_{\rm line} = \langle\sigma v \rangle_{\gamma\gamma} + 1/2 \langle\sigma v \rangle_{\gamma Z}$, versus the DM mass compared to the theoretical line cross section computed at NLO for three canonical DM SU(2) models: the Wino (light green)~\cite{Baumgart:2018yed}, the Higgsino (light blue)~\cite{Beneke:2022eci}, and Quintuplet (light red)~\cite{Baumgart:2023pwn}. For each model, the mass for which the correct relic abundance~\cite{Planck:2018vyg} is obtained from a thermal cosmology is shown as the vertical band~\cite{Montanari:2022buj}. The corresponding thermal cross sections are highlighted for the Wino, Higgsino, and Quintuplet as green, blue and red dots, respectively.
The observed upper limits are shown for all the Milky Way DM density models adopted in this work.
}
\label{fig:Upperlimits}
\end{figure*}
\section{Results}
\label{sec:results}
No significant excess compatible with a DM annihilation line signal is found in any of the 
spectral and spatial bins with respect to the expected background.
We therefore derive 95\% C.L. upper limits on 
$\langle \sigma v \rangle_{\rm line}$, for DM masses from 300 GeV up to 70 TeV. The mean expected upper limits and 68 and 95\% containment bands are computed to quantify
the consistency of observed limits with the null hypothesis.
The left panel of Fig.~\ref{fig:Upperlimits} shows the observed and expected upper limits on $\langle \sigma v \rangle_{\rm line}$, as a function of the DM mass for the baseline Einasto DM profile with $r_{\odot}$ = 8.178~kpc \cite{GRAVITY:2018ofz}. To illustrate the impact of systematic uncertainties, we also show the mean expected upper limits without including the systematic uncertainty nuisance parameter in the TS. The strongest limit reaches $\langle \sigma v \rangle_{\rm line} = 9.8 \times 10^{-29}$ cm$^3$s$^{-1}$ for a DM mass of $\sim$440 GeV. For DM masses of 1 and 10 TeV, the limits reach $2.3$ $\times$ $10^{-28}$ and $2.4 \times$ $10^{-27}$ cm$^3$s$^{-1}$, respectively. 
The systematic uncertainty on the energy scale of the count distributions is 10\%, which similarly affects the measured and expected energy count distributions and would lead to a 10\% overall shift of the limits along the DM mass axis~\cite{HESS:2022ygk}. Degrading the energy resolution artificially to $\sigma_E/E$ = 15\% worsens the limits by $\sim 1.4$ for a 1 TeV DM mass. Following the extensive study presented in Ref.~\cite{HESS:2022ygk}, no other systematic uncertainties need to be considered for this analysis and dataset.

The theoretical cross sections for the Wino~\cite{Baumgart:2018yed}, Higgsino~\cite{Beneke:2022eci} and Quintuplet~\cite{Baumgart:2023pwn} DM models computed at the next-to-leading order (NLO) are plotted on the right panel of Fig.~\ref{fig:Upperlimits}. The mass values to thermally produce the DM density measured today are shown for the three models as well, with the associated uncertainty. The thermal Wino and Quintuplet DM models are excluded for any of the adopted DM profiles. Our limits with the baseline Einasto profile probe for the first time the long-hunted thermal cross section value for Higgsino DM. 
If an alternative production mechanism to thermal production is assumed, which changes the relevant DM mass, the Wino and Quintuplet models are excluded up to about 10 TeV, and 25 TeV, respectively. 
We also show the impact of adopting the alternative MW DM density models
Thermal Wino and Quintuplet are excluded for all the MW DM density models considered here. 
With the Auriga DM profile, the theoretical line cross section of the Higgsino is probed up to about 10 TeV. At 500 GeV, changing $r_{\odot}$ = 8.178~kpc to $r_{\odot}$ = 8.5~kpc as used in previous literature \cite{MAGIC:2022acl, HESS:2022ygk}, improves the limits with the Einasto profile by about 10\%. If the NFW profile is adopted, the limits degrade by a factor $\sim 1.2$ compared to the baseline Einasto. For a cored profile such as the cNFW, the limits strongly worsen by a factor $\sim 5.8$. Using MW DM density models from hydrodynamic simulations of MW-like galaxies improves the limits by a factor $\sim 1.4$ and $\sim 2.9$ with FIRE-2 and Auriga, respectively.

\begin{figure}[htbp!]
\includegraphics[width=0.45\textwidth]{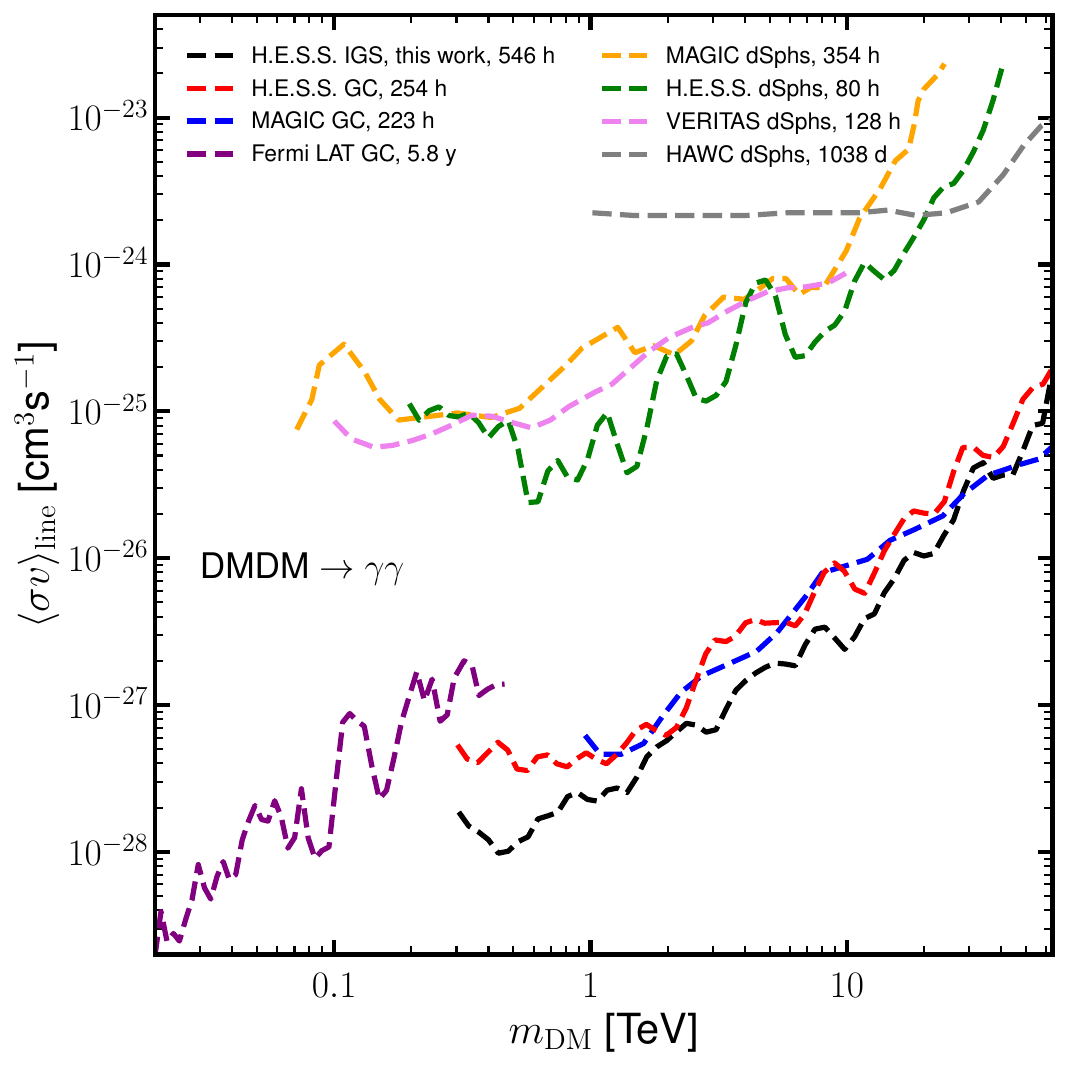}
\caption{Current constraints on the line cross section $\langle \sigma v \rangle_{\rm line}$ versus the DM mass $m_{\rm DM}$  including previous H.E.S.S. limits from 254 h of observations of the GC~\cite{Abdallah:2018qtu} (red line), the limits from 223 h of GC observations with MAGIC~\cite{MAGIC:2022acl} (blue line), and the limits from 5.8 y of observations of the GC with the
Fermi satellite~\cite{Fermi-LAT:2015kyq} (purple line). The limits from dwarf spheroidal galaxy observations with HAWC~\cite{HAWC:2019jvm} (grey line), H.E.S.S.~\cite{HESS:2020zwn} (green line), MAGIC~\cite{MAGIC:2021mog} (yellow line) and VERITAS~\cite{VERITAS:2017tif} (pink line) are also
plotted. } 
\label{fig:SummaryPlot}
\end{figure}

Figure~\ref{fig:SummaryPlot} compares our observed upper limits with those obtained with several instruments and from observations of different regions of the sky. Our analysis significantly improves the previous constraints on the line cross section from 300 GeV up to $\sim 30$ TeV. Large photon statistics from longer observation time and optimized pointing positions, the exploitation of the full H.E.S.S. five-telescope array, and the application of run-wise simulations for the IRFs~\cite{Holler:2020duc} drive the sensitivity improvement. For a 1 TeV DM mass, the limit is improved by a factor of $\sim 2.1$ with respect to the previous H.E.S.S. results. The constraints obtained in this work are the strongest so far in probing multi-TeV DM models. More details are provided in Ref.~\cite{supplement_more_details}.

\section{Summary}
In this Letter, we report the latest results from the search for annihilating DM line signals with the H.E.S.S. Inner Galaxy Survey of the inner halo of the Milky Way.
The present dataset amounts to 546 h of total live time achieved in 6 years of data taking. The absence of significant line excess provides the strongest constraints on the velocity-weighted annihilation cross section into two photons for Majorana WIMPs in the 300 GeV to 30 TeV mass range. This work provides strong constraints on electroweak DM candidates such as the Wino, Higgsino, and Quintuplet, which arise as the most 
minimal and elegant extensions of the Standard Model. The observed limits with the baseline Einasto profile reach $\langle \sigma v \rangle_{\rm line} = 2.3$ $\times$ $10^{-28}$, $6.5 \times 10^{-28}$, and $5.8 \times 10^{-27}$ cm$^3$s$^{-1}$ at 1, 2.9 and 13.8 TeV, respectively. Thermal Wino and Quintuplet models were already excluded.
However, our constraints 
are the first to reach the thermal mass for Higgsino DM, one of the most compelling DM models that escapes current and near-future detection prospects in direct detection and collider experiments. 
Its thermal cross-section value is probed for various Milky Way DM density models.

VHE observations of the central region of the Milky Way with Southern-hemisphere IACT arrays such as H.E.S.S. are crucial to probe and study in-depth canonical WIMP models that are beyond the reach of direct detection and collider searches. H.E.S.S. observations of the GC region, under the most favourable conditions among currently operating IACTs, are unique for exploring the yet-uncharted parameter space of multi-TeV DM models and provide crucial insights into the TeV WIMP paradigm. The IGS program carried out with H.E.S.S. is an important legacy and paves the way for future Southern-site observations of the GC region with the Cherenkov Telescope Array Observatory.

\section{Acknowledgements}
The support of the Namibian authorities and of the University of Namibia in facilitating the construction and operation of H.E.S.S. is gratefully acknowledged, as is the support by the German Ministry for Education and Research (BMBF), the Max Planck Society, the German Research Foundation (DFG), the Helmholtz Association, the Alexander von Humboldt Foundation, the French Ministry of Higher Education, Research and Innovation, the Centre National de la Recherche Scientifique (CNRS/IN2P3 and CNRS/INSU), the Commissariat \`a l'\'energie atomique et aux \'energies alternatives (CEA), the U.K. Science and Technology Facilities Council (STFC), the Knut and Alice Wallenberg Foundation, the National Science Centre, Poland grant no. 2016/22/M/ST9/00382, the South African Department of Science and Technology and National Research Foundation, the University of Namibia, the National Commission on Research, Science \& Technology of Namibia (NCRST), the Austrian Federal Ministry of Education, Science and Research and the Austrian Science Fund (FWF), the Australian Research Council (ARC), the Japan Society for the Promotion of Science and by the University of Amsterdam.

We appreciate the excellent work of the technical support staff in Berlin, Zeuthen, Heidelberg, Palaiseau, Paris, Saclay, T\"ubingen and in Namibia in the construction and operation of the equipment. This work benefitted from services provided by the H.E.S.S. Virtual Organisation, supported by the national resource providers of the EGI Federation.

\newpage
\newpage

\bibliography{bibl}

\appendix
\newpage
\setcounter{equation}{0}
\setcounter{figure}{0}
\widetext
\begin{center}
{\bf \large \large Supplemental Material: Search for gamma-ray spectral lines from dark matter annihilation in the H.E.S.S. Inner Galaxy Survey}
\end{center}

\input{supplement}

\end{document}

%% file: authors_prl.tex
\author{H.E.S.S. Collaboration}
\noaffiliation

\author{F.~Aharonian}
\affiliation{Astronomy \& Astrophysics Section, School of Cosmic Physics, Dublin Institute for Advanced Studies, DIAS Dunsink Observatory, Dublin D15 XR2R, Ireland}
\affiliation{Max-Planck-Institut für Kernphysik, P.O. Box 103980, D 69029 Heidelberg, Germany}

\author{H.~Ashkar}
\affiliation{Laboratoire Leprince-Ringuet, École Polytechnique, CNRS, Institut Polytechnique de Paris, F-91128 Palaiseau, France}

\author{V.~Barbosa~Martins}
\affiliation{Deutsches Elektronen-Synchrotron DESY, Platanenallee 6, 15738 Zeuthen, Germany}

\author{R.~Batzofin}
\affiliation{Institut für Physik und Astronomie, Universität Potsdam, Karl-Liebknecht-Strasse 24/25, D 14476 Potsdam, Germany}

\author{Y.~Becherini}
\affiliation{Université Paris Cité, CNRS, Astroparticule et Cosmologie, F-75013 Paris, France}

\author{D.~Berge}
\affiliation{Deutsches Elektronen-Synchrotron DESY, Platanenallee 6, 15738 Zeuthen, Germany}
\affiliation{Institut für Physik, Humboldt-Universität zu Berlin, Newtonstr. 15, D 12489 Berlin, Germany}

\author{K.~Bernl\"ohr}
\affiliation{Max-Planck-Institut für Kernphysik, P.O. Box 103980, D 69029 Heidelberg, Germany}

\author{M.~B\"ottcher}
\affiliation{Centre for Space Research, North-West University, Potchefstroom 2520, South Africa}

\author{C.~Boisson}
\affiliation{LUX, Observatoire de Paris, Université PSL, CNRS, Sorbonne Université, 5 Pl. Jules Janssen, 92190 Meudon, France}

% ...

%% file: supplement.tex


\section{Observations and data analysis}

A deep and wide observation program is conducted by the H.E.S.S. collaboration to survey the Galactic Center (GC) region. In particular, H.E.S.S. observed the region between 2014 and 2020 with the full 5-telescope array, for a live time of 546 hours, with zenith angles below 45$^\circ$, and an average of 18$^\circ$~\cite{HESS:2022ygk}. The pointing positions for the observations were defined on a grid with Galactic latitudes $b$ of $0.8^\circ$, $1.6^\circ$ and $3.2^\circ$ above the Galactic plane, and Galactic longitudes $l$ from $-1.8^\circ$ up to $1.8^\circ$, with $1.2^\circ$ spacing. For the row at $b=0.8^\circ$, two additional pointings were added at $l = -3.0 ^\circ$ and $l = 3.0^\circ$. The 546 hours dataset was obtained after selecting good-quality data, according to the standard quality selection procedure~\cite{Aharonian:2006pe,deNaurois:2009ud}. 

The gamma-ray events were selected and reconstructed by fitting the shower images to semi-analytical models~\cite{deNaurois:2009ud}. This technique yields an angular resolution of $0.06^\circ$ (68\% containment radius) and an energy resolution of 10\% above  300 GeV. The reconstructed gamma-like event position is required to be within a distance of 2.5$^\circ$ from the pointing position. The line signal is searched for in regions of interest (ROIs) defined as concentric annuli centred on the GC position, as extensively explained in Ref.~\cite{HESS:2022ygk}. The centre of the Milky Way is a complex environment harbouring numerous regions emitting VHE gamma rays. A conservative set of masks is applied to avoid the challenging modelling of VHE emission and gamma-ray contamination from known astrophysical sources in the field of view. The inner ROI radii range between 0.5$^\circ$ and 2.9$^\circ$, with a width of 0.1$^\circ$ each. The residual gamma-ray background is measured for each annulus on a run-by-run basis. The regions used for the background measurement (OFF) are built symmetrically to the ROIs, used for signal measurement (ON). The parts of the sky falling inside the masks are equally removed from the ON and OFF regions, ensuring that the same solid angle is obtained for the two. Given the sufficient distance between the ON and the OFF regions for each run, a significant difference in the expected DM signal, \textit{i.e.}, the J-factors, in the signal and background measurement regions is always obtained when considering cuspy DM profiles. Potential unaccounted gamma-ray emissions, which may be present outside the masks, are considered part of the measured excess, making, therefore, the derived upper limits conservative \cite{HESS:2022ygk}.
The number of events measured with H.E.S.S. is correlated with the zenith angle of the observation. Therefore, a gradient is expected across the telescope's field of view. The procedure explained in Ref.~\cite{HESS:2022ygk} is followed to mitigate this effect.

\section{Test-statistics and upper limit derivation}
\label{sec:ts}
The derivation of the upper limits on
$\langle \sigma v\rangle_{\rm line}$
is carried out via a log-likelihood ratio test statistic (TS) to test the hypothesis of a DM signal against the null hypothesis in the data, assuming a positive searched signal, {\it i.e.}, $\langle  \sigma v \rangle >$ 0. The TS is defined as~\cite{2011EPJC711554C}: 
\begin{equation}
TS = 
\begin{cases}
- 2\, \rm ln \frac{\mathcal{L}(N^{\rm S}(\langle \sigma v \rangle), \widehat{\widehat {N^{\rm B}}}(\langle \sigma v \rangle)
)}{\mathcal{L}(0,\widehat{\widehat{N^{\rm B}(0)}}
)} \,  & N^{\rm S}(\widehat{\langle \sigma v \rangle}) < 0\\
- 2\, \rm ln \frac{\mathcal{L}(N^{\rm S}(\langle \sigma v \rangle), \widehat{\widehat {N^{\rm B}}}(\langle \sigma v \rangle)
)}{\mathcal{L}(N^{\rm S}(\widehat{\langle \sigma v \rangle}),\widehat{N^{\rm B}}
)} \,  & 0\leq N^{\rm S}(\widehat{\langle \sigma v \rangle}) \leq N^{\rm S}(\langle \sigma v \rangle)\\
0 & N^{\rm S}(\widehat{\langle \sigma v \rangle}) > N^{\rm S}(\langle \sigma v \rangle) \, .
\end{cases}
\label{eq:TS}
\end{equation}

$N^{\rm S}$ corresponds to the sum of the number of gamma rays expected from DM annihilation for the observational run $k$ $N^{\rm S}_{\rm k}$, over all runs $k$. In a region of solid angle $\Delta\Omega$ with a J-factor $J(\Delta\Omega$), the self-annihilation of Majorana DM particles of mass $m_{\rm DM}$ with a thermally-averaged annihilation cross section
$\langle \sigma v \rangle_{\rm line}$, is expected to produce $N^{\rm S}_{\rm k}$ events given by:
\begin{equation}
 N^{\text{S}}_{\text k}(\langle \sigma v \rangle) =  \frac{\langle \sigma v \rangle J(\Delta\Omega)}{8\pi m_{\rm DM}^2} T_{\rm{obs},k} \int_{E_{\rm th}}^{m_{\rm DM}} \int^{\infty}_{0}  
 \frac{dN^{\rm line}_{\gamma}}{dE_{\gamma}}(E_{\gamma}) \: R(E_{\gamma}, E'_{\gamma}) \: A_{\rm eff, k}(E_{\gamma}) 
 \:  dE_{\gamma} \: dE'_{\gamma}\, .
\end{equation}
The finite energy resolution is expressed by $R(E_{\gamma}, E'_{\gamma})$, which relates the energy detected $E'_{\gamma}$ to the true energy $E_{\gamma}$ of the events. $A_{\rm eff, k}(E_{\gamma})$ and $T_{\rm obs, k}$ are the energy-dependent acceptance and the observation time for the run $k$, respectively.
For each run $k$ of given observational and instrumental detector conditions, the energy-dependent acceptance is computed according to the semi-analytical shower model template technique using standard selection cuts~\cite{deNaurois:2009ud} according to the run-wise simulation scheme~\cite{Holler:2020duc}. 
For each run, the spatial response of the instrument is encoded in the acceptance term, which depends on the angular distance between the reconstructed event position and the pointing position of the run $k$. The energy resolution is well described by a Gaussian function of $\sigma/E$ of 10\% above 200~GeV~\cite{deNaurois:2009ud}. 

$\widehat{\widehat {N^{\rm B}_{\rm ij}}}$ is derived via a conditional maximization by solving $\partial \mathcal{L}/\partial N^{\rm B}_{\rm ij} = 0$. This corresponds to the conditional maximum likelihood estimator of $\mathcal{L}$, {\it i.e.}, the value of $N^{\rm B}_{\rm ij}$
that maximizes 
$\mathcal{L}$ for the given $N^{\rm S}_{\rm ij}(\langle \sigma v \rangle)$. $N^{\rm S}_{\rm ij}(\widehat{\langle \sigma v \rangle})$ and $\widehat{N^{\rm B}_{\rm ij}}$ are derived using an unconditional maximization, {\it i.e.}, they are the maximum likelihood estimators of $\mathcal{L}$.

No significant VHE gamma-ray excess compatible with the searched DM signal is found in any of the ROI; therefore, upper limits on the thermally-averaged velocity-weighted annihilation cross section $\langle  \sigma v \rangle_{\rm line}$ for a set of DM masses $m_{\rm DM}$ can be derived with the TS. One-sided upper limits at 95\% C. L. are computed by requiring a TS value of 2.71. This assumes that the TS follows a $\chi^2$ distribution, as expected in the high statistics limit, with one degree of freedom.

The data analysis utilises the expected spectral and spatial characteristics of the searched DM signal with respect to residual background. Therefore, the 
total likelihood function is given by the product of Poisson likelihood functions over the spatial and energy bins, \textit{i.e.}, $\mathcal{L} = \prod_{\rm ij}\mathcal{L}_{\rm ij}$.
Considering the two-body DM annihilation taking place almost at rest, the DM-induced line gamma-ray spectrum in the final state provides a smoking-gun signature against the much smoother power-law-like spectrum of the residual background. The spatial morphology of the expected DM signal, following the spatial J-factor profile, provides additional discriminating power, given the spatially independent morphology of the residual background.

\section{Dark matter halo models for the Milky Way and J-factors}
\label{sec:jfactor}
An important challenge in computing upper limits on the dark matter (DM) annihilation cross section is posed by the determination of the DM distribution in the inner Milky Way (MW). In addition to the mass modelling approach based on direct kinematic measurements of the gravitational potential, one can use cosmological simulations of structure formation to derive the DM density profile for the MW.

If one relies on DM-only simulations for MW-like galaxies~\cite{Springel:2008by, Diemand:2008in, 10.1111/j.1745-3933.2009.00699.x}, the MW DM halo can be well parameterised by the Navarro-Frenk-White (NFW)~\cite{Navarro:1996gj} or Einasto~\cite{Springel:2008by} profiles, predicting cuspy DM distributions. However, the complexity in the simulation is dramaticaly increased when baryonic physics and feedback processes are included~\cite{10.1093/mnras/stw145,10.1093/mnras/sty1690,2019MNRAS.490.4877P}. Such processes can, for instance, lead to the development of kpc-sized cores in the MW center~\cite{Mollitor:2014ara,Chan:2015tna}. Nevertheless, baryonic feedback can create a contraction of the inner profile, which enhances the DM density at small radii~\cite{Hopkins:2017ycn,McKeown:2021sob}. These several reasons prevent a firm assessment of the DM distribution in the innermost region of the MW, which translates into a systematic uncertainty that affects the indirect detection of DM signals. To bracket this uncertainty, we have considered five different MW DM models for the computation of the J-factors.

We first consider the Einasto DM density profile as a baseline for direct comparison with previous results, with the parametrisation extracted from~\cite{Springel:2008by}:
\begin{equation}
\label{eq:Einastoprofile}
\rho_\text{Einasto}  \propto \exp\left[-\frac{2}{\alpha} \left( \left(\frac{r}{r_s}\right)^\alpha - 1 \right)\right]\!,
\end{equation}
with $\alpha = 0.17$ and $r_s=20$ kpc. The Einasto profile is normalized to the local DM density $\rho_\odot$ such that $\rho_{{\rm Einasto}}(r_\odot) = \rho_\odot = 0.39$ GeV/cm$^3$~\cite{Catena:2009mf}. Limits obtained with Einasto are shown for $r_\odot = 8.178$ kpc \cite{GRAVITY:2018ofz}--considered as baseline parameterisation, and $r_\odot = 8.5$ kpc \cite{Ghez:2008ms}, respectively. An improvement of the limits by about 10\% is observed using $r_\odot = 8.5$ kpc. 
Adding baryonic physics and feedback processes in DM-only simulations can lead to modified DM profiles, including a contraction of the inner DM distribution, possibly showing a core-like behaviour, significantly enhancing the DM density in the inner few degrees~\cite{Hopkins:2017ycn,Board:2021bwj,McKeown:2021sob}. Such a behaviour is also observed in mass modelling approaches~\cite{Portail:2016vei,Cautun:2019eaf,Lin:2019yux}.
Following the work in Ref.~\cite{Cautun:2019eaf}, we consider DM density profiles derived from mass modelling, making use of a determination of the Milky Way mass profile obtained with Gaia DR2 measurements of the rotation curve, and an in-depth modelling of the baryonic components in the GC region. The resulting profile can be reasonably parameterised by NFW profiles, expressed by: 
\begin{equation}
\label{eq:NFWprofile}
\rho_{{\rm NFW}(r)} \propto \frac{1}{(r/r_s)(1+r/r_s)^2}.
\end{equation}
with the parameters reported in Tab.~\ref{tab:profiles}. However, a deviation from the standard NFW parametrization is observed~\cite{Cautun:2019eaf}. The inferred DM distribution shows evidence of being contracted by the presence of baryons~\cite{Cautun:2019eaf}. The derived model provided by the authors is non-parametric, however we refer to it as contracted NFW (cNFW). We emphasize, however, that the cNFW does not follow the parametrization given in Eq.~\ref{eq:NFWprofile}. The approach followed in Ref.~\cite{Cautun:2019eaf} does not allow for a measurement of the distribution within the inner 1 kpc of the inner Galaxy. For this profile, we therefore define a core in the inner region such that $\rho_{{\rm cNFW}}(r) = \rho_{{\rm cNFW}}(r_c)$ for $r \le r_c = 1~{\rm kpc}$ \cite{Montanari:2022buj}. However, this is a conservative case, because the DM density may continue to increase towards the GC, rather than taking on a core. Note that DM cores larger than several kpc are disfavored by stellar measurements~\cite{Hooper:2016ggc}. Adopting the NFW and cNFW profiles, weakens the limits on the DM cross section by a factor $\sim 1.3$ and $\sim 6.5$, respectively. 

From the results of hydrodynamical simulations, we adopt the DM models of MW from Ref.~\cite{McKeown:2021sob}, extracted from FIRE-2 zoom-in simulations of MW-like galaxies. These simulations include radiative heating and cooling for gas, stellar feedback from OB stars, type Ia and type II supernovae, radiation pressure, and star formation effects. As the authors of Ref.~\cite{McKeown:2021sob} considered several simulations, we extracted the geometrical mean of all the J-factor profiles to compute our limits. The minimum spatial resolution of the simulations is $\sim 400$ pc, therefore, we have extrapolated the J-factor profile to the GC using a linear approximation as done, e.g., in Refs.~\cite{Montanari:2022buj, PhysRevD.108.083027}. Such an extrapolation creates an almost flat density profile below the inner $\sim 400$ pc~\cite{Montanari:2022buj}. Finally, we consider the geometrical mean of the profiles extracted from the Auriga simulations of MW-like galaxies~\cite{Hussein:2025xwm}. As for the FIRE-2 simulations, the Auriga profiles are modelled following adiabatic contraction. However, FIRE-2 has stronger baryonic feedback, leading to more suppressed DM density in the inner few degrees. A much less aggressive suppression is observed in the Auriga profile, resulting in larger DM density close to the GC~\cite{Hussein:2025xwm}.

We normalise the different DM density profiles $\rho(r)$ to the local DM density $\rho_\odot$ at the Solar distance $r_\odot$ such that $\rho(r_\odot) = \rho_\odot$. The Einasto and NFW DM profiles are normalised such that $\rho$($r_\odot$) = $\rho_\odot = 0.39$ GeV cm$^{-3}$~\cite{Catena:2009mf}. The value used for the local DM density for each DM model is extracted from the corresponding references and reported in Tab.~\ref{tab:profiles}. Nevertheless, the DM density at the solar location is subject to uncertainties, and the values used here may be updated as more accurate determinations of $\rho_\odot$ are obtained. Following Refs.~\cite{Read:2014qva,Zyla:2020zbs,deSalas:2020hbh}, determinations of $\rho_\odot$ are in the range (0.2 $-$ 0.6) GeV cm$^{-3}$. Note that recent determinations using GRAVITY and GAIA measurements suggest (0.4 $-$ 0.8) GeV cm$^{-3}$~\cite{Benito:2019ngh}.
As the local density could be more precisely determined by new observational measurements, changes to $\rho_\odot$ can be propagated by rescaling the DM signal, obtained from DM distributions inferred from CDM-only simulations, by ($\rho_\odot$/$\rho_{\odot, {\rm profile}})^2$, where $\rho_{\odot, {\rm profile}}$ is what is reported in Tab.~\ref{tab:profiles}. However, this rescaling may not be valid for distributions obtained from hydrodynamical simulations.
\begin{table}[h]
 \centering
 \begin{tabular}{l c c c c c c}
 \hline
 \hline
MW DM models                      & $\rho_{\odot}$  & $r_{\rm s}$ & $r_{\rm c}$  & $\alpha_{\rm s}$ & $r_{\odot}$  & Reference \\
                     & (GeVcm$^{-3}$) &  (kpc) &  (kpc) &  &  (kpc) &  \\
  \hline
Einasto & 0.39 & 20   &  /    & 0.17 & 8.178 \cite{GRAVITY:2018ofz} & \cite{Springel:2008by}\\
Einasto & 0.39 & 20   &  /    & 0.17 & 8.5 \cite{Ghez:2008ms} & \cite{Springel:2008by}\\
NFW 	& 0.32 & 15.5 &  /    & /    & 8.0 & \cite{Cautun:2019eaf} \\
cNFW    & 0.34 & 23.8 &  1    & /    & 8.0 & \cite{Cautun:2019eaf} \\
FIRE-2  & 0.38 &  /   &  /    &  /   & 8.3 & \cite{McKeown:2021sob} \\
Auriga  & 0.30 &  /   &  /    &  /   & 8.0 & \cite{Hussein:2025xwm} \\
 \hline
 \hline
 \end{tabular}
 \caption{Parameters of the mass density profiles used for the spherical Milky Way DM models. 
 \label{tab:profiles}}
 \end{table}

\begin{figure}[!ht]
 \centering
 \includegraphics[width=0.4\textwidth]{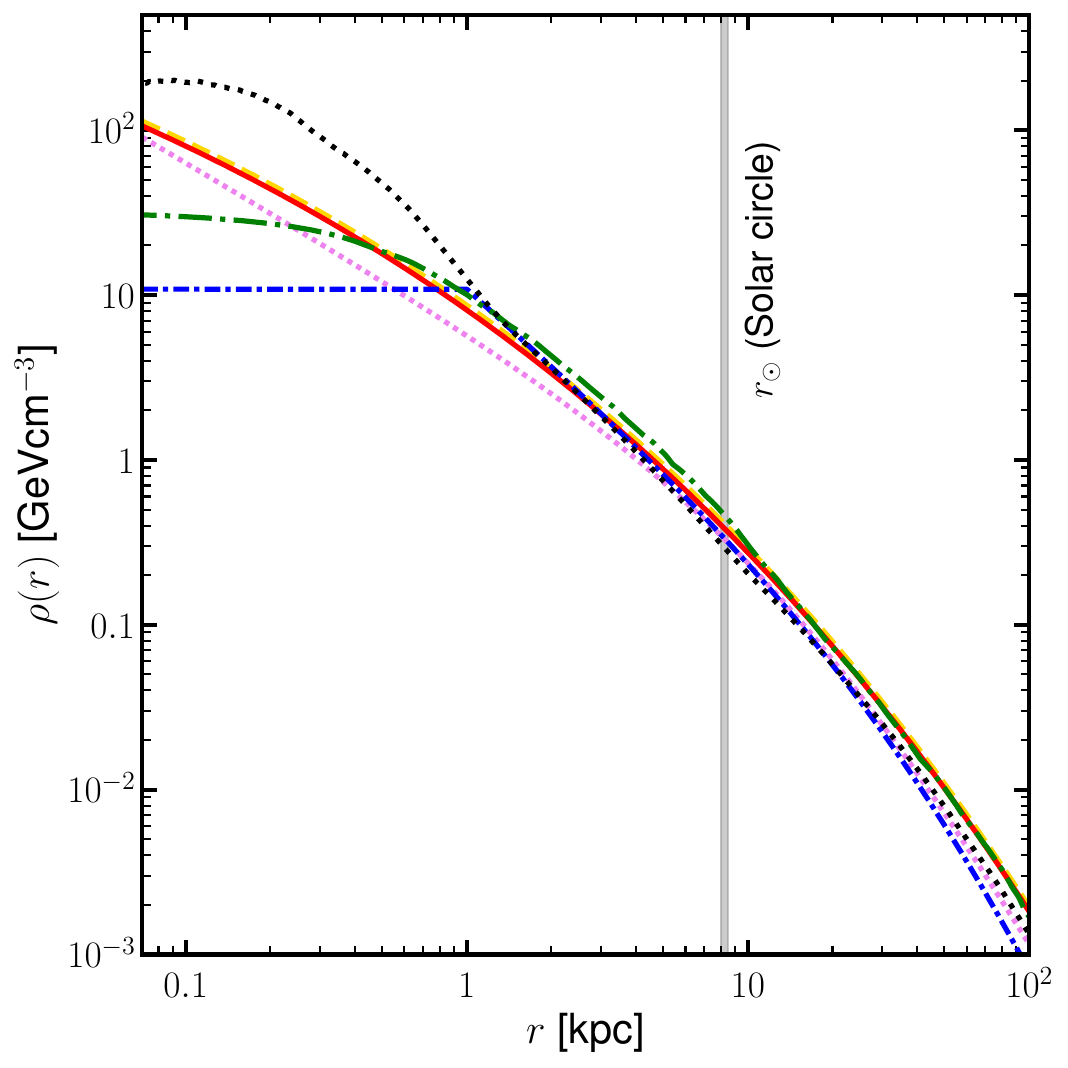}
 \includegraphics[width=0.4\textwidth]{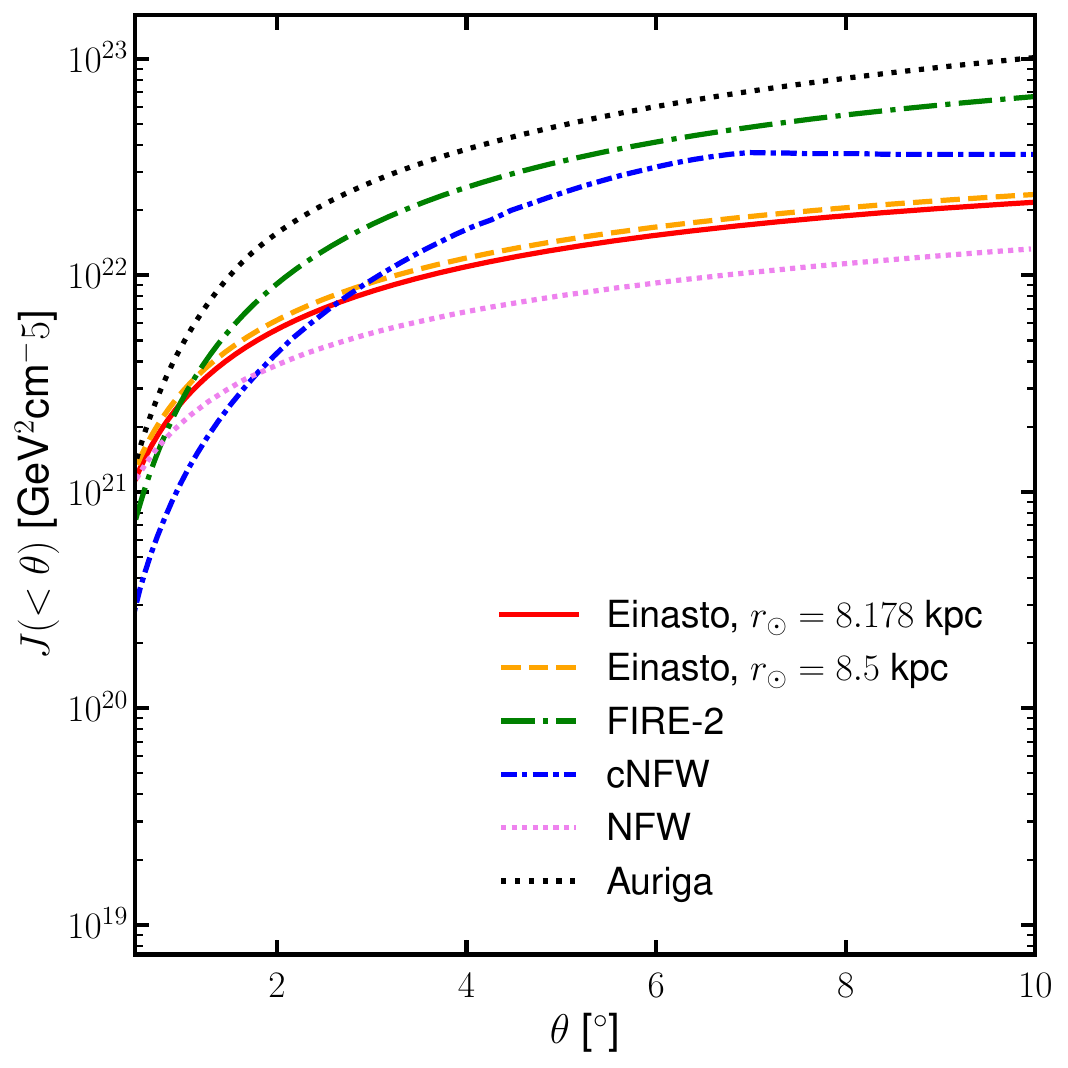}
 \caption{\textit{Left panel:} DM mass density profile $\rho$ (GeVcm$^{-3}$) as a function of the distance $r$ (kpc) from the Galactic Centre distance for the DM halo models of the Milk Way discussed here. The Einasto profile is plotted assuming $r_\odot$ = 8.5 kpc (orange dashed line) as in Refs.~\cite{HESS:2022ygk,MAGIC:2022acl,Ghez:2008ms} and $r_\odot$ = 8.178 kpc (red solid line) following Ref.~\cite{GRAVITY:2018ofz}. The NFW and cNFW profiles (purple densely dotted and blue dash-dotted lines) are extracted from Refs.~\cite{Cautun:2019eaf,PhysRevD.108.083027}. The non-parametric geometric mean profiles extracted from the hydrodynamical simulations FIRE-2~\cite{McKeown:2021sob} and Auriga~\cite{Hussein:2025xwm} (green long-dash-dotted and black dotted lines), respectively, are also displayed. The profiles are normalized according to $\rho(r_\odot)$ displayed in Tab.~\ref{tab:profiles}. 
 The vertical grey band encompasses the values used for the solar distance $r_{\odot}$ adopted for the computation of the different J-factor profiles used in this work. \textit{Right panel:} Integrated J-factors (GeV$^2$cm$^{-5}$) as a function of the angular distance (deg.) from the Galactic Centre for the DM mass density profiles on the left panel. Since exclusion regions mask a part of the solid angle in our analysis, the effective J-factors are reduced compared to the ones displayed here.}
 \label{fig:DMdensityprofiles}
 \end{figure}

The DM density profiles (GeV cm$^{-3}$), as a function of the distance from the GC in kpc, are displayed for all the profiles adopted in this work in the left panel of Fig.~\ref{fig:DMdensityprofiles}. The integrated J-factors (GeV$^2$cm$^{-5}$), as a function of the angular distance $\theta$ from the GC (in $^\circ$), are displayed for all the profiles used in this work in the right panel of Fig.~\ref{fig:DMdensityprofiles}. Both panels show profiles down to the inner radius of the first annulus of the region of interest (ROI) for the analysis. The integrated J-factors (GeV$^2$cm$^{-5}$) in each ROI used in this work are reported in Tab.~\ref{tab:jfactors} for all the profiles. In particular, the J-factor integrated over all the ROI rings for the Einasto profile is 2.5$\times$10$^{22}$ GeV$^2$cm$^{-5}$. Applying as distance to the centre of the Milky Way $r_\odot = 8.178$ kpc \cite{GRAVITY:2018ofz}, decreases the integrated J-factor to 2.3$\times$10$^{22}$ GeV$^2$cm$^{-5}$. For the NFW profile, the integrated J-factor results to 1.0$\times$10$^{22}$ GeV$^2$cm$^{-5}$. Instead, modelling DM with the contracted cNFW results in an integrated J-factor of 2.3$\times$10$^{22}$ GeV$^2$cm$^{-5}$. Finally, the integrated J-factors for the FIRE-2 and Auriga profiles are 5.1$\times$10$^{22}$ GeV$^2$cm$^{-5}$ and 7.9$\times$10$^{22}$ GeV$^2$cm$^{-5}$, respectively.

\begin{table}[ht!]
\footnotesize
\centering
\begin{tabular}{c|c|c|c|c|c|c|c|c|c}
\hline
\hline
$i^{\rm th}$ ROI & Inner radius & Outer radius & Solid angle $\Delta\Omega$&\multicolumn{6}{c}{$J$-factor $J(\Delta\Omega)$}  \\
 &  [deg.] &  [deg.] & [10$^{-4}$ sr] &\multicolumn{6}{c}{[10$^{20}$ GeV$^2$cm$^{-5}$]}  \\
\hline
& & & & Einasto & Einasto  &  NFW & cNFW & FIRE-2 & Auriga \\
& & & & $r_\odot$ = 8.5 kpc & $r_\odot$ = 8.178 kpc  &   &  &  &  \\
\hline
1  & 0.5 & 0.6 & 1.05 & 9.4  & 8.5 & 4.2 & 2.9  & 8.5  & 15.2 \\
2  & 0.6 & 0.7 & 1.24 & 9.9  & 8.9 & 4.3 & 3.5  & 10.2 & 18.1 \\
3  & 0.7 & 0.8 & 1.44 & 10.0 & 9.0 & 4.3 & 4.1  & 11.6 & 20.6 \\
4  & 0.8 & 0.9 & 1.63 & 10.2 & 9.3 & 4.2 & 4.7  & 13.3 & 23.4 \\
5  & 0.9 & 1.0 & 1.82 & 10.3 & 9.4 & 4.2 & 5.1  & 14.4 & 25.4 \\
6  & 1.0 & 1.1 & 2.01 & 10.3 & 9.4 & 4.2 & 5.7  & 15.6 & 27.6 \\
7  & 1.1 & 1.2 & 2.20 & 10.4 & 9.4 & 4.2 & 6.2  & 16.8 & 29.8 \\
8  & 1.2 & 1.3 & 2.39 & 10.4 & 9.4 & 4.1 & 6.7  & 17.9 & 31.7 \\
9  & 1.3 & 1.4 & 2.58 & 10.4 & 9.4 & 4.2 & 7.3  & 18.9 & 33.3 \\
10 & 1.4 & 1.5 & 2.78 & 10.5 & 9.6 & 4.1 & 7.9  & 20.1 & 35.2 \\
11 & 1.5 & 1.6 & 2.97 & 10.3 & 9.3 & 4.1 & 8.3  & 20.4 & 35.4 \\
12 & 1.6 & 1.7 & 3.16 & 10.2 & 9.4 & 4.0 & 8.9  & 21.4 & 35.9 \\
13 & 1.7 & 1.8 & 3.35 & 10.2 & 9.3 & 4.0 & 9.4  & 21.9 & 35.2 \\
14 & 1.8 & 1.9 & 3.54 & 10.2 & 9.3 & 3.9 & 9.9  & 22.7 & 35.2 \\
15 & 1.9 & 2.0 & 3.73 & 10.0 & 9.2 & 3.9 & 10.4 & 23.1 & 34.7 \\
16 & 2.0 & 2.1 & 3.92 & 10.0 & 9.2 & 3.9 & 10.9 & 23.7 & 34.9 \\
17 & 2.1 & 2.2 & 4.11 & 9.8  & 9.0 & 3.9 & 11.4 & 23.9 & 34.6 \\
18 & 2.2 & 2.3 & 4.31 & 9.8  & 8.9 & 3.9 & 11.9 & 24.4 & 34.9 \\
19 & 2.3 & 2.4 & 4.50 & 9.7  & 8.9 & 3.9 & 12.4 & 24.6 & 34.9 \\
20 & 2.4 & 2.5 & 4.69 & 9.6  & 8.8 & 3.9 & 12.9 & 24.9 & 34.9 \\
21 & 2.5 & 2.6 & 4.88 & 9.6  & 8.8 & 3.9 & 13.4 & 25.3 & 35.2 \\
22 & 2.6 & 2.7 & 5.07 & 9.5  & 8.7 & 3.8 & 13.9 & 25.5 & 35.3 \\
23 & 2.7 & 2.8 & 5.26 & 9.3  & 8.5 & 3.8 & 14.3 & 25.5 & 35.2 \\
24 & 2.8 & 2.9 & 5.45 & 9.3  & 8.5 & 3.8 & 14.9 & 25.9 & 35.6 \\
25 & 2.9 & 3.0 & 5.64 & 9.1  & 8.4 & 3.7 & 15.2 & 25.7 & 35.2 \\
\hline
\hline
\end{tabular}
\caption{J-factor values in each of the 25 annuli for the MW DM models considered in this work.  The first column provides the ROI number with its inner and outer radii in the second and third columns, respectively, and its solid angle in the fourth column. The fifth, sixth, seventh, eighth, ninth and tenth columns display the total J-factor values in the ROI for the two Einasto profiles -- with $r_\odot = 8.5$ kpc~\cite{HESS:2022ygk,MAGIC:2022acl,Ghez:2008ms} and $r_\odot = 8.178$ kpc~\cite{Ghez:2008ms}, the NFW~\cite{Cautun:2019eaf} and cNFW~\cite{Cautun:2019eaf,Montanari:2022buj} profiles, and the FIRE-2~\cite{McKeown:2021sob} and Auriga~\cite{Hussein:2025xwm} profiles, respectively. Since exclusion regions mask a part of the solid angle in our analysis, the effective J-factors are reduced compared to the ones displayed here.
\label{tab:jfactors}}
\end{table}